\documentclass[12pt]{article}
\usepackage{amssymb}

\usepackage{amsmath}

\newtheorem{theorem}{Theorem}
\newtheorem{acknowledgement}[theorem]{Acknowledgement}

\begin{document}

\author{D. Galetti \\
Instituto de F\'{i}sica Te\'{o}rica\\
S\~{a}o Paulo State University - UNESP\\
Rua Pamplona 145\\
01405 - 900 S\~{a}o Paulo, SP, Brazil}
\date{}
\title{Quantum description of spin tunneling in magnetic molecules: a new view}
\maketitle

\begin{abstract}
Starting from a phenomenological Hamiltonian originally written in terms of
angular momentum operators we derive a new quantum angle-based Hamiltonian
that allows for a discussion on the quantum spin tunneling. The study of the
applicability of the present approach, carried out in calculations with a
soluble quasi-spin model, shows that we are allowed to use our method in the
description of physical systems such as the Mn12-acetate molecule, as well
as the octanuclear iron cluster, Fe8, in a reliable way. With the present
description the interpretation of the spin tunneling is seen to be direct,
the spectra and energy barriers of those systems are obtained, and it is
shown that they agree with the experimental ones.
\end{abstract}

PACS: 75.45+j, 75.60Jp\\
Keywords: Spin tunneling, Lipkin model, Mn12-acetate, Fe8 cluster

\bigskip 
Corresponding author \newline
D. Galetti \newline
Instituto de F\'{i}sica Te\'{o}rica - UNESP \newline
Phone: (011) 3177 9001 \newline
Fax: (011) 3177 9080 \newline
e-mail: galetti@ift.unesp.br \newpage

\section{Introduction}

Besides the well-known examples of tunneling processes that permeate the
literature in molecular and nuclear physics \cite{gamow,gurney,bell}, in
recent years it has been shown that quantum tunneling also appears in as
broad physical situations as magnetic moments in spin glass systems, single
magnetic impurities in a crystal field, a single domain, a ferromagnetic, or
antiferromagnetic, domain wall, and also in spin tunneling in molecules \cite
{tejada,chudgunt,stamp}, emphasizing the importance of the angle/angular
momentum degree of freedom in these cases. This quantum effect occuring in
such mesoscopic, and even in macroscopic systems points to the necessity of
a discussion on how such tunneling can be realized and controlled \cite
{takagi}.

In what concerns the case of spin tunneling, or the quantum tunneling of
magnetization, QTM, the theoretical approaches have pointed to some
different ways of treating this problem. In particular, the spin tunneling
was first treated by van Hemmen and S\"{u}to \cite{van}, and by Enz and
Schilling \cite{enz}. From different perspectives, various approaches were
introduced: by using a generalization of the usual WKB method adapted to
spin systems \cite{scharf,zaslav,van2}, or by using Feyman's path integral
treatment of quantum mechanics \cite{Schill}, and also by using coherent
states \cite{ulianov}. On the other hand, from the experimental point of
view, great advances have been made since the synthesis of the Mn12-acetate
molecule in 1980 \cite{lis}. Among the properties of this molecule it
emerged an indication of spin tunneling that has strongly motivated great
efforts in characterizing and elucidating the pure quantum contribution to
that phenomenon \cite{sessoli,novak,paulsen,friedman,thomas}. The discovery
of the same indications of spin tunneling in the octanuclear iron cluster,
Fe8 \cite{barra,sangregorio}, has thus reinforced the proposed idea of pure
quantum spin tunneling, and formal treatments involving phenomenological
Hamiltonians were then widely proposed and discussed \cite{caneschi}.

In any case, the angular momentum operators obeying the standard commutation
relations 
\begin{equation*}
\left[ J_{i},J_{j}\right] =i\varepsilon _{ijk}J_{k}
\end{equation*}
constitute the fundamental blocks from which the starting Hamiltonian
describing the spin system is constructed. In general, a Hamiltonian is
proposed based on experimental considerations on the symmetries of the
physical system, and also under the \ assumption of low temperature such
that the pure quantum effects dominate. In this sense, a general Hamiltonian
encompassing the molecule symmetries plus possible paralel and tranverse
magnetic fields may be written as 
\begin{equation}
H=H_{0}+C_{1}J_{z}+C_{2}J_{x},  \label{int1}
\end{equation}
being $H_{0}$ the molecule Hamiltonian, and the constants $C_{i}$ measure
the intensity of the additional external magnetic fields. This description
clearly separates and takes into account the molecule total angular momentum
degree of freedom as the dominant one and, as far as no other degree of
freedom is directly and explicitly involved, it is considered a sound
description of the system. As such, all of its kinematical content can be
described in terms of combinations of the set of $2j+1$ states generated by
all the projections of the total angular momentum operator $J$, which obeys
the eigenvalue equation 
\begin{equation*}
J^{2}|jm\rangle =\hbar ^{2}j(j+1)|jm\rangle \;.
\end{equation*}
It is then clear that those Hamiltonians can then be diagonalized within the
state space defined by $\left\{ |jm\rangle \right\} $ whose dimension is
given by the total angular momentum $J$. At the same time, as it is well
known, since the Hamiltonian is written in terms of the generators of that
algebra, it commutes with the Casimir operator thus giving at least one more
constant of the motion - besides the energy; in fact, all the symmetries
present in the Hamiltonian must also play a precise role, giving rise to
characteristic features in the energy spectrum.

In this paper we want to extend the content of previous publications \cite
{garu,gapili} that proposed to construct and present the main features of a
potential function, described in terms of an angle variable, which can be
used to picture in a more intuitive way the spin tunneling. In those works,
we have shown how to obtain such a potential and how pure quantum tunneling
shows up as a transmition through an energy barrier. In the present
contribution we show how to extract a new Hamiltonian which also includes a
kinetic term - not only a potential energy function - also described in
terms of an angle variable, from a phenomenological Hamiltonian, Eq.(\ref
{int1}), that can be used to study the spin tunneling.

To construct such a Hamiltonian, as for the kinematics, we use the $su(2)$
coherent states \cite{perelomov}, since they constitute an overcomplete
basis of states, and also because they present the desired properties for
the study of the systems of interest in the semiclassical limit. At the same
time they also include quantum effects such as zero point energies
associated with the corresponding angle and angular momentum variables. We
also use the Generator Coordinate Method (GCM) of wide use in nuclear and
molecular physics\cite{griwhee,wong} mainly because it has been shown to be
a useful tool when one deals with collective behavior in many-body systems,
that is the case in the spin systems of interest where the tunneling degree
of freedom has a suggested collective character.

From the operational point of view, the GCM is used so that an energy
surface is generated that embodies off-diagonal matrix elements that account
for the quantum corrections to the diagonal contribution which, by its turn,
is basically dominated by the semiclassical contribution. The present
approach proposes to take into account those off-diagonal contributions by
calculating the moments of that energy distribution; the zeroth-order moment
is associated with the potential function, while the second-order moment is
associated with an inertia function: together they constitute our
approximate Hamiltonian. It is clear that the higher order terms can also be
obtained so that the complete series gives rise to a new representation of
the starting phenomenological Hamiltonian. A truncated series up to the
second term is then an approximate description of the system, and its
validity must necessarily be discussed. It is precisely the use of the $%
su(2) $ coherent states that makes this study feasible since our moment
expansion of the energy surface is then akin to the $1/N$ expansion
discussed in the literature \cite{yaffe}, which also makes explicit use of
the coherent states. In the present context, it is the total spin of the
molecule, $j$, the quantum number that determines how many terms we must
keep in the moment expansion series in order to get a reliable Hamiltonian.

It is worth noticing that the present approach is based on quantum grounds
from the beginning in such a way that it does not need to be quantized in
any form. Therefore, it has a different perspective from the approaches that
start from a classical Hamiltonian and then quantize it. Furthermore, what
is important to keep in mind in the present context is that the associated
pair of quantum operators related to the rotation degree of freedom must be
treated properly\cite{suss,carnieto,lynch}.

This paper is organized as follows. In Section II we introduce the $su(2)$
coherent states in the discussion of the semiclassical approach for a
general phenomenological Hamiltonian, and show how some preliminary results
are obtained which suggest that quantum corrections may be important in some
cases. The GCM is presented in Section III where we show how the energy and
overlap kernels of the method are obtained for the general phenomenological
Hamiltonian, and the new general angle-based Hamiltonian is obtained. In
this section we also carry out the calculations for the soluble Lipkin
quasi-spin model \cite{lipkin,marshalek} since it is a valuable testing
ground for systems of the kind we are interested in, and also for
establishing a reliable truncation criterion for the moment series in terms
of the value of the spin. The result that we obtain guarantees the
application, in section IV, of the second-order approximate Hamiltonian to
the description of two interesting molecules studied in the literature,
namely, the Mn12-acetate, and the Fe8 cluster, and we show that we reproduce
the experimental results to a good degree of accuracy. The spin tunneling
phenomenon occuring in these molecules is qualitatively discussed in this
context, and the main aspects of the potential barrier are discussed. The
role of the effective inertia function in the tunneling process is also
stressed. Finally the conclusions are presented in section V.

\section{A semiclassical approach}

As a starting point to get a semiclassical approach for describing spin
systems we may consider the $su(2)$ coherent states defined in a general
form as \cite{perelomov} 
\begin{equation*}
|jz\rangle =\left( 1+\left| z\right| ^{2}\right) ^{-j}e^{zJ_{+}}|j,-j\rangle
,
\end{equation*}
or also as 
\begin{equation*}
|jz\rangle =\left( 1+\left| z\right| ^{2}\right) ^{-j}\sum_{m=-j}^{j}\binom{%
2j}{j+m}z^{j+m}|jm\rangle ,
\end{equation*}
where $z$ is a complex variable given by 
\begin{equation*}
z=\tan \frac{\alpha }{2}e^{-i\xi },
\end{equation*}
and $j$ characterizes the angular momentum state multiplet. These states are
nonorthogonal and normalized, and furthermore satisfy an overcompleteness
relation 
\begin{equation*}
\int |jz\rangle \langle jz|d\mu \left( z\right) =1,
\end{equation*}
which allows us to use them as an useful state basis.

Moreover, these states are a very interesting starting point when
semiclassical aspects of the physical system must be discussed, as has been
already discussed in the literature \cite{yaffe}.

Let us consider now a spin Hamiltonian of the type that, for simplicity,
does not include transversal magnetic fields 
\begin{equation}
\emph{H}=AJ_{z}+BJ_{z}^{2}+G(J_{+}^{2}+J_{-}^{2})  \label{hbasica}
\end{equation}
written in terms of the basic operators obeying the $su(2)$ algebra 
\begin{equation*}
\left[ J_{z},J_{\pm }\right] =\pm J_{\pm }
\end{equation*}
and 
\begin{equation*}
\left[ J_{+},J_{-}\right] =2J_{z}.
\end{equation*}
It is clearly seen that by properly choosing the parameters of that
Hamiltonian we end up with particular expressions associated with physical
systems we may want to study, for instance if we choose $B=0$ we have the
Lipkin quasi-spin solvable model \cite{lipkin} 
\begin{equation*}
\emph{H}_{L}=AJ_{z}+G\left( J_{+}^{2}+J_{-}^{2}\right) \;.
\end{equation*}
By choosing $G=0$ we get the Mn-12 acetate model, where a paralel magnetic
fields is present \cite{tejada}, namely 
\begin{equation*}
\emph{H}_{Ac}=AJ_{z}+BJ_{z}^{2}\;,
\end{equation*}
and we also see that by choosing $A=0$ we have a model Hamiltonian for the
octanuclear iron cluster, [Fe$_{8}$O$_{2}\left( \text{OH}\right) _{12\text{ }%
}\left( \text{tacn}\right) _{6}$]$^{8+}$, or just Fe8 \cite
{barra,sangregorio}, in the absence of external magnetic fields 
\begin{equation*}
\emph{H}_{Fe}=BJ_{z}^{2}+G\left( J_{+}^{2}+J_{-}^{2}\right) \text{.}
\end{equation*}

We then recall that a description based on an angle representation of these
models, out of that general Hamiltonian, can be obtained by directly
calculating its respective normalized matrix elements with the proposed
coherent states. To this end, we collect the basic results 
\begin{equation*}
\frac{\langle jz|J_{z}|jz\rangle }{\langle jz|jz\rangle }=-j\cos \alpha \;,
\end{equation*}

\begin{equation*}
\frac{\langle jz|J_{z}^{2}|jz\rangle }{\langle jz|jz\rangle }=j^{2}-j\left(
j-\frac{1}{2}\right) \sin ^{2}\alpha \;,
\end{equation*}
\begin{equation*}
\frac{\langle jz|J_{+}^{2}|jz\rangle }{\langle jz|jz\rangle }=2j\left(
2j-1\right) \sin ^{2}\frac{\alpha }{2}\cos ^{2}\frac{\alpha }{2}e^{2i\xi
},\;\;
\end{equation*}
and 
\begin{equation*}
\frac{\langle jz|J_{-}^{2}|jz\rangle }{\langle jz|jz\rangle }=\left[ \frac{%
\langle jz|J_{+}^{2}|jz\rangle }{\langle jz|jz\rangle }\right] ^{\dagger }.
\end{equation*}
Using these results we get the energy surface associated with the
Hamiltonian, namely, 
\begin{eqnarray}
\mathcal{H}\left( \alpha ,\xi \right) &=&-Aj\cos \alpha +B\left[
j^{2}-j\left( j-\frac{1}{2}\right) \sin ^{2}\alpha \right] +
\label{hamisemi} \\
&&2G2j\left( 2j-1\right) \sin ^{2}\frac{\alpha }{2}\cos ^{2}\frac{\alpha }{2}%
\cos 2\xi \;.  \notag
\end{eqnarray}
At this point, we see that we can determine the value of $\xi $ that
minimizes the energy surface, leading to an energy curve in the variable $%
\alpha $.

Let us analyse this expression in a particular case. It is evident that, if
we choose the parameters so that we describe the Mn12-acetate molecule in
the absence of paralel and transverse magnetic fields, i.e., $A=G=0$ (the
energy surface in this case is $\xi $-independent), the energy curve 
\begin{equation*}
\mathcal{H}\left( \alpha \right) =B\left[ j^{2}-j\left( j-\frac{1}{2}\right)
\sin ^{2}\alpha \right]
\end{equation*}
presents minima at $\alpha =0,\pi $ for $B<0$. For the values $B/k_{B}=-0.6$ 
$K$, where $k_{B}$ is the Boltzmann constant, and $j=S=10$, as found in the
literature \cite{sessoli}, we obtain the energy minima at $E_{\min }=-60.0$ $%
K$. This is, however, precisely the ground state energy value obtained by
diagonalizing the Hamiltonian associated with the corresponding model, Eq.(%
\ref{hbasica}), within the $|j(=S)m\rangle $ states basis. This result thus
indicates that such a semiclassical approach does not embody quantum content
enough to describe such system because, as we see, the bottom of the energy
curve coincides with the ground state energy. This fact shows that the
quantum correlations related to the rotation degree of freedom may be
important and, furthermore, it is not taken into account in this approach.

\section{Formalism}

Let us now go one step further in order to construct an energy surface that
describes the spin system and at the same time that encompasses more quantum
correlations than the previous procedure does. In this sense the Generator
Coordinate Method (GCM) \cite{griwhee} extends that approach because it also
embodies off-diagonal elements, i.e., if we start from the $su(2)$ coherent
states we have to calculate the so-called energy and overlap kernels, as
already discussed\cite{garu,gapili} 
\begin{equation*}
\emph{K}\left( \alpha ,\alpha ^{\prime };\xi \right) =\langle jz^{\prime }|%
\emph{H}|jz\rangle =\langle j\alpha ^{\prime }\xi |\emph{H}|j\alpha \xi
\rangle =\langle \alpha ^{\prime }|\emph{H}|\alpha \rangle
\end{equation*}
\begin{equation*}
\emph{N}\left( \alpha ,\alpha ^{\prime }\right) =\langle jz^{\prime
}|jz\rangle =\langle j\alpha ^{\prime }\xi |j\alpha \xi \rangle =\langle
\alpha ^{\prime }|\alpha \rangle ,
\end{equation*}
respectively. It must be observed that hereafter we will only consider $%
\alpha $ as our generator coordinate. The variable $\xi $ will be properly
chosen at the beginning in each case so as to minimize the energy surface as
pointed out before.

In the same way as we did in the last section, the basic matrix elements of
the operators are collected

\begin{equation*}
\langle jz^{\prime }|J_{z}|jz\rangle =-j\cos ^{2j-1}\left( \frac{\alpha
^{\prime }-\alpha }{2}\right) \cos \left( \frac{\alpha ^{\prime }+\alpha }{2}%
\right) \;,
\end{equation*}

\begin{equation*}
\langle jz^{\prime }|J_{z}^{2}|jz\rangle =\frac{j}{2}\cos ^{2j}\left( \frac{%
\alpha ^{\prime }-\alpha }{2}\right) +\frac{j}{2}\left( 2j-1\right) \cos
^{2j-2}\left( \frac{\alpha ^{\prime }-\alpha }{2}\right) \cos ^{2}\left( 
\frac{\alpha ^{\prime }+\alpha }{2}\right) \;,
\end{equation*}
and 
\begin{equation*}
\langle jz^{\prime }|J_{+}^{2}+J_{-}^{2}|jz\rangle =2j\left( 2j-1\right)
\cos ^{2j-2}\left( \frac{\alpha ^{\prime }-\alpha }{2}\right) \times
\end{equation*}
\begin{equation*}
\left( e^{2i\xi }\sin ^{2}\frac{\alpha ^{\prime }}{2}\cos ^{2}\frac{\alpha }{%
2}+e^{-2i\xi }\sin ^{2}\frac{\alpha }{2}\cos ^{2}\frac{\alpha ^{\prime }}{2}%
\right) .
\end{equation*}
Therefore, the GCM energy and overlap kernels for the general spin
Hamiltonian are respectively 
\begin{equation*}
\emph{K}\left( \alpha ,\alpha ^{\prime };\xi \right) =-Aj\cos ^{2j-1}\left( 
\frac{\alpha ^{\prime }-\alpha }{2}\right) \cos \left( \frac{\alpha ^{\prime
}+\alpha }{2}\right) +B\frac{j}{2}\cos ^{2j}\left( \frac{\alpha ^{\prime
}-\alpha }{2}\right) +
\end{equation*}
\begin{equation*}
B\frac{j}{2}\left( 2j-1\right) \cos ^{2j-2}\left( \frac{\alpha ^{\prime
}-\alpha }{2}\right) \cos ^{2}\left( \frac{\alpha ^{\prime }+\alpha }{2}%
\right) +
\end{equation*}
\begin{equation}
2Gj\left( 2j-1\right) \cos ^{2j-2}\left( \frac{\alpha ^{\prime }-\alpha }{2}%
\right) \left( e^{2i\xi }\sin ^{2}\frac{\alpha ^{\prime }}{2}\cos ^{2}\frac{%
\alpha }{2}+e^{-2i\xi }\sin ^{2}\frac{\alpha }{2}\cos ^{2}\frac{\alpha
^{\prime }}{2}\right) ,  \label{hgcm}
\end{equation}
and 
\begin{equation}
N\left( \alpha ,\alpha ^{\prime }\right) =\cos ^{2j}\left( \frac{\alpha
^{\prime }-\alpha }{2}\right) .  \label{ogcm}
\end{equation}
The contribution of the off-diagonal matrix elements is immediately
recognized in these expressions, and it can be easily verified that they are
less and less important as $j$ increases, and that the semiclassical results
are thus reobtained for $\alpha ^{\prime }=\alpha $, or when $j\rightarrow
\infty $. Thus, the off-diagonal contributions may play an essential role in
the description of spin systems with small and medium values of $j$ since
they carry additional quantum information when compared with the
semiclassical approach.

Instead of solving the GCM integral equation with these kernels, as usually
is done\cite{griwhee}, we will show in what follows how a spin Hamiltonian
-- written in terms of an angle variable -- can be extracted from the GCM
kernels.

Once we have the GCM overlap and energy kernels, we will use a new
represention for describing the spin system\cite{pipa}. Following the scheme
already presented before\cite{garu,gapili}, which makes use of Fourier
transforms, we get the new matrix representation for the spin system 
\begin{equation*}
H_{nn^{\prime }}=-\frac{A}{2}\frac{S\left( j,n,n^{\prime }\right) }{\Gamma
\left( \frac{2j+1}{2}+\frac{n+n^{\prime }}{2}\right) \Gamma \left( \frac{2j+1%
}{2}-\frac{n+n^{\prime }}{2}\right) }\left( \delta _{n^{\prime },n+1}+\delta
_{n^{\prime },n-1}\right) +
\end{equation*}
\begin{equation*}
B\frac{j}{2}\frac{S\left( j,n,n^{\prime }\right) }{\Gamma \left( j+\frac{%
n+n^{\prime }}{2}+1\right) \Gamma \left( j-\frac{n+n^{\prime }}{2}+1\right) }%
\delta _{n^{\prime },n}+
\end{equation*}
\begin{equation*}
\frac{B}{4}\frac{S\left( j,n,n^{\prime }\right) }{\Gamma \left( j+\frac{%
n+n^{\prime }}{2}\right) \Gamma \left( j-\frac{n+n^{\prime }}{2}\right) }%
\left[ 2\delta _{n^{\prime },n}+\left( \delta _{n^{\prime },n+2}+\delta
_{n^{\prime },n-2}\right) \right] +
\end{equation*}
\begin{equation*}
\frac{G}{2}\frac{S\left( j,n,n^{\prime }\right) }{\Gamma \left( j+\frac{%
n+n^{\prime }}{2}\right) \Gamma \left( j-\frac{n+n^{\prime }}{2}\right) }%
\left[ 6\delta _{n^{\prime },n}-\left( \delta _{n^{\prime },n+2}+\delta
_{n^{\prime },n-2}\right) \right] +
\end{equation*}
\begin{equation}
Gj\left( 2j-1\right) \frac{S\left( j,n,n^{\prime }\right) }{\Gamma \left( j+%
\frac{n+n^{\prime }}{2}+1\right) \Gamma \left( j-\frac{n+n^{\prime }}{2}%
+1\right) }\delta _{n^{\prime },n}\;,  \label{exata}
\end{equation}
where 
\begin{equation*}
S\left( j,n,n^{\prime }\right) =\sqrt{\Gamma \left( j+n+1\right) \Gamma
\left( j-n+1\right) \Gamma \left( j+n^{\prime }+1\right) \Gamma \left(
j-n^{\prime }+1\right) }.
\end{equation*}
The diagonalization of this Hamiltonian matrix clearly gives the same
results as those obtained by directly diagonalizing the matrix generated by
the starting Hamiltonian, Eq.(\ref{hbasica}), using the angular momentum $%
|jm\rangle $ states basis. This result is expected because this new
representation is obtained by the use of unitary transformations acting on
the original one. This also assures us that this new representation totally
preserves the content of the original representation and that it is not an
approximate version of the energy surface.

This new representation has also the virtue of allowing a further
manipulation that is an essential step in order to write a Hamiltonian in
terms of an angle variable for the spin system. Namely, to this end we
perform new Fourier transformations, followed by a discrete version of the
continuous Weyl-Wigner transformation \cite{degroot}. Again, using the
results of Refs. \cite{garu,gapili}, we can get an expression for the energy
surface - from the GCM energy kernel - in terms of a new pair of angle
variables, namely $\beta $ and $\beta ^{\prime }$. In fact, we introduce the
new variables $\varphi =\frac{\alpha +\alpha ^{\prime }}{2},\;\theta =\alpha
^{\prime }-\alpha ,\;u=\beta -\beta ^{\prime }$ and $\phi =\frac{\beta
+\beta ^{\prime }}{2}$, and also $l=n+n^{\prime }$ and $k=n-n^{\prime }$ in
order to transform the GCM energy kernel into the new energy surface. The
summations over $l$ that appear are restricted so that they run over
even/odd values if $k$ is even/odd.

Since we already have the GCM energy kernel, Eq.(\ref{hgcm}), we will
consider hereafter the particular case of no transverse magnetic field, $C=0$%
, for simplicity, and therefore we rewrite it as 
\begin{equation*}
H^{\left( GCM\right) }\left( \theta ,\varphi \right) =\langle \varphi +\frac{%
\theta }{2}|H|\varphi -\frac{\theta }{2}\rangle =
\end{equation*}
\begin{equation*}
\left[ B\frac{j}{2}+\left| G\right| j\left( 2j-1\right) \right] \cos
^{2j}\left( \frac{\theta }{2}\right) -2\left| G\right| j\left( 2j-1\right)
\cos ^{2j-2}\left( \frac{\theta }{2}\right) -
\end{equation*}
\begin{equation*}
Aj\cos ^{2j-1}\left( \frac{\theta }{2}\right) \cos \varphi +j\left(
2j-1\right) \left( \frac{B}{2}+\left| G\right| \right) \cos ^{2}\varphi ,
\end{equation*}
where the choices for the variable $\xi $ were already carried out. After a
direct but tedious calculation we obtain the general expression, 
\begin{equation*}
\emph{H}\left( u,\phi \right) =\frac{\left[ B\frac{j}{2}+\left| G\right|
j\left( 2j-1\right) \right] }{2\pi }\sum_{l=-2j}^{2j}e^{i\frac{l}{2}u}+\frac{%
\frac{B}{2}-3\left| G\right| }{2\pi }\sum_{l=-2j}^{2j}e^{i\frac{l}{2}%
u}\left( j^{2}-\frac{l^{2}}{4}\right) -
\end{equation*}
\begin{equation*}
\frac{A}{2\pi }\cos \phi \sum_{l=-2j}^{2j}e^{i\frac{l}{2}u}\sqrt{\left( j+%
\frac{l}{2}+\frac{1}{2}\right) \left( j-\frac{l}{2}+\frac{1}{2}\right) }+
\end{equation*}
\begin{equation}
\frac{\frac{B}{2}+\left| G\right| }{2\pi }\sum_{l=-2j}^{2j}e^{i\frac{l}{2}%
u}\cos 2\phi \sqrt{\left( j+\frac{l}{2}+\frac{1}{2}\right) \left( j+\frac{l}{%
2}\right) \left( j-\frac{l}{2}+\frac{1}{2}\right) \left( j-\frac{l}{2}%
\right) }.  \label{superficiegeral}
\end{equation}
So far no approximation has been made and the above expression contains
exactly the same physical content as the original phenomenological
Hamiltonian.

\subsection{Discussion}

Before carrying out the calculations in order to obtain the moments of the
above shown energy surface, we must recall some aspects of the Weyl-Wigner
transformation techniques developed for continuous degrees of freedom \cite
{gasa}. It can be readily recognized that the power series expansion in the
momentum $p$ that appears in the defining expression of the Weyl-Wigner
transform 
\begin{equation*}
H\left( q,p\right) =\int e^{ip\left( x-x^{\prime }\right) }H\left( \frac{%
x+x^{\prime }}{2},x-x^{\prime }\right) d\left( x-x^{\prime }\right) ,
\end{equation*}
where $\frac{x+x^{\prime }}{2}=q$, and $H\left( q,p\right) $ is the quantum
phase space representative of the Hamiltonian operator, can be rewritten in
such a form that the new coefficients are the moments of the energy surface
with respect to the difference of the original variables, namely

\begin{equation*}
M_{n}\left( \frac{x+x^{\prime }}{2}\right) =\int H\left( \frac{x+x^{\prime }%
}{2},x-x^{\prime }\right) \left( x-x^{\prime }\right) ^{n}d\left(
x-x^{\prime }\right) ,\;\;n=0,2,4,...
\end{equation*}
is the n-th moment of the energy surface in the variable $x-x^{\prime }$.
Obviously we expect that only the even moments do not vanish because
otherwise the system would violate time reversal symmetry, or, in other
words, terms would appear in the Hamiltonian that would be proportional to
odd powers of the momentum. In fact, it is the even parity of the GCM
kernels that warrants us a time reversal invariant Hamiltonian.

In the continuous case, by using this procedure one ends up with a series
the first two terms of which are dependent on the coordinate variable only,
and proportional to the squared momentum respectively. In some cases, the
proportionality coefficient in the second term may result in a function of
the coordinate variable or, in other words, an effective mass may come out.
Therefore, these first two terms can be seen as the potential and kinetic
energy terms of an associated effective Hamiltonian. When these two terms
dominate the series expansion, i.e., higher order moments can be neglected%
\emph{\ }-- they may even be completely absent in certain cases -- then the
approximate two-terms effective Hamiltonian can be useful to describe the
physical system we started with.

Now, the same can be seen to be valid in the present case of spin degrees of
freedom if we consider the proper angular function in respect to which the
moments of the energy surface must be taken. Since the angle is not the
direct variable to be used here since it does not obey a standard
commutation relation with the angular momentum operator \cite
{suss,carnieto,lynch} -- in contrast with the canonical case of ordinary
coordinate and momentum discussed above -- the moments of the energy surface
must be calculated with a periodic function of the angle. Thus, we perform
the calculation with respect to $\sin (u)$ \cite{carnieto}, where $u=\beta
-\beta ^{\prime }$, as already defined. The zeroth order moment will give us
the potential function in terms of the angle variable, as has been already
presented in the past \cite{garu,gapili}. Since the zeroth order is given by 
\begin{equation*}
\emph{V}\left( \phi \right) =\sum_{u=-\frac{2j}{2j+1}\pi }^{\frac{2j}{2j+1}%
\pi }\emph{H}\left( u,\phi \right) \Delta u,
\end{equation*}
where $\Delta u=\frac{2\pi }{2j+1}$, we end up with 
\begin{equation*}
\emph{V}\left( \phi \right) =\frac{j\left[ B+2\left| G\right| \left(
2j-1\right) \right] }{2}+\frac{B-6\left| G\right| }{2}j^{2}-
\end{equation*}
\begin{equation*}
\frac{A\cos \phi }{2j+1}\sqrt{j\left( j+1\right) }\sum_{n,k=-j}^{j}e^{\frac{%
2\pi i}{N}}e^{i\pi \frac{k}{N}}\sqrt{1-\frac{n\left( n+1\right) }{j\left(
j+1\right) }}\;+
\end{equation*}
\begin{equation}
\frac{B+2\left| G\right| }{2}j\left( j+1\right) \cos 2\phi ,  \label{pot}
\end{equation}
where $N=2j+1$ is the angular momentum state space dimension.

By its turn, the second moment will give us the expression associated with
the inertia function. In other words, 
\begin{equation*}
\emph{I}\left( \phi \right) =\sum_{u=-\frac{2j}{2j+1}\pi }^{\frac{2j}{2j+1}%
\pi }\emph{H}\left( u,\phi \right) \sin ^{2}u\;\Delta u
\end{equation*}
leads to 
\begin{equation*}
\emph{I}\left( \phi \right) =\left( B-6\left| G\right| \right) +\frac{A\cos
\phi }{4j(j+1)}\sqrt{j\left( j+1\right) }\times
\end{equation*}
\begin{equation*}
\sum_{n,k=-j}^{j}e^{i\pi \frac{k}{N}}\left[ e^{2\pi i\frac{k}{N}\left(
n+2\right) }+e^{2\pi i\frac{k}{N}\left( n-2\right) }-2e^{2\pi i\frac{k}{N}n}%
\right] \sqrt{1-\frac{n\left( n+1\right) }{j\left( j+1\right) }}\;-
\end{equation*}
\begin{equation*}
\frac{B+2\left| G\right| }{4}j\left( j+1\right) \cos 2\phi \left[ \sqrt{1-%
\frac{6}{j\left( j+1\right) }}\sqrt{1-\frac{2}{j\left( j+1\right) }}-1\right]
.
\end{equation*}
It is then a direct matter to see that 
\begin{equation*}
\emph{I}\left( \phi \right) =-\frac{1}{\emph{M}\left( \phi \right) },
\end{equation*}
where $\emph{M}\left( \phi \right) $ is the corresponding effective ''mass''
function associated with the spin system.

We now propose that these first two moments must suffice to write a
Hamiltonian that can, to an acceptable degree of accuracy, describe systems
consisting of a suitably great number of spins. In this form, in this
approximation the Hamiltonian is written as 
\begin{equation*}
\emph{H}\left( \phi \right) =-\frac{d}{d\phi }\left( \frac{1}{2\emph{M}%
\left( \phi \right) }\frac{d}{d\phi }\right) +\emph{V}\left( \phi \right) ,
\end{equation*}
and we straightforwardly get the eigenvalues $E_{k}$ as well as the
eigenfunctions $\psi _{k}\left( \phi \right) $ by solving the
Schr\"{o}dinger equation 
\begin{equation*}
\emph{H}\left( \phi \right) \psi _{k}\left( \phi \right) =E_{k}\psi
_{k}\left( \phi \right)
\end{equation*}
by a Fourier analysis. As such, the spin number for which the validity of
this proposal is assured can be estimated by introducing a model Hamiltonian
and comparing the results obtained from this Schr\"{o}dinger equation with
the exact ones, obtained from a direct diagonalization of that model
Hamiltonian in a $|jm\rangle $ basis, for a great range of values of the
spin quantum number $j$. In what follows we will discuss the validity of
that approximation using the Lipkin model\cite{lipkin}.

As already pointed out before, the Lipkin quasi-spin model is singled out by
assuming the parameters: $A=\varepsilon $, $B=0$, and $G=\frac{V}{2}=-\frac{%
\varepsilon \chi }{2\left( N_{s}-1\right) }$ , where $N_{s}=2j$, so that the
GCM kernels read 
\begin{equation*}
\frac{\emph{H}_{L}\left( \theta ,\varphi \right) }{\varepsilon }=-\frac{N_{s}%
}{2}\{\cos ^{N_{s}-1}\left( \frac{\theta }{2}\right) \cos \varphi +\frac{%
\chi }{2}\cos ^{N_{s}-2}\left( \frac{\theta }{2}\right) \times
\end{equation*}
\begin{equation*}
\left[ \left( 1+\sin ^{2}\varphi \right) -\cos ^{2}\left( \frac{\theta }{2}%
\right) \right] \},
\end{equation*}
and 
\begin{equation*}
N_{L}\left( \theta ,\varphi \right) =\cos ^{N_{s}}\left( \frac{\theta }{2}%
\right)
\end{equation*}
respectively, if we measure energy in unities of $\varepsilon $.

We obtain the potential function directly from Eq.(\ref{pot}), which gives 
\begin{equation}
\emph{V}_{L}\left( \phi \right) =-\frac{\chi j\left( j+1\right) }{2j-1}\sin
^{2}\phi -\frac{\sqrt{j\left( j+1\right) }}{N}\cos \phi \times
\label{exapot}
\end{equation}
\begin{equation*}
\sum_{n,k=-j}^{j}e^{2\pi i\frac{k}{N}\left( n+\frac{1}{2}\right) }\sqrt{1-%
\frac{n\left( n+1\right) }{j\left( j+1\right) }}.
\end{equation*}
In the same form we get the expression related to the inertia function 
\begin{equation*}
\emph{I}_{L}\left( \phi \right) =-\frac{3\chi }{2j-1}+\frac{\sqrt{j\left(
j+1\right) }}{4N}\cos \phi \times
\end{equation*}
\begin{equation*}
\sum_{n,k=-j}^{j}e^{i\pi \frac{k}{N}}\left[ e^{2\pi i\frac{k}{N}\left(
n+2\right) }+e^{2\pi i\frac{k}{N}\left( n-2\right) }-2e^{2\pi i\frac{k}{N}n}%
\right] \sqrt{1-\frac{n\left( n+1\right) }{j\left( j+1\right) }}-
\end{equation*}
\begin{equation}
\frac{\chi j\left( j+1\right) }{4\left( 2j-1\right) }\cos 2\phi \left[ \sqrt{%
1-\frac{6}{j\left( j+1\right) }}\sqrt{1-\frac{2}{j\left( j+1\right) }}-1%
\right] .  \label{exainer}
\end{equation}

With these expressions we can discuss now the approximations proposed
before. First, if the number of quasi-spins is such that $N_{s}\gg 1$, the
summations over $k$ can be substituted by integrals and we obtain the
expressions 
\begin{equation*}
\emph{V}_{L}\left( \phi \right) =-\frac{\chi \left( N_{s}+3\right) }{4}\sin
^{2}\phi -\sqrt{j\left( j+1\right) }\cos \phi \frac{2}{\pi }\sum_{n=-j}^{j}%
\frac{\left( -1\right) ^{n}}{2n+1}\sqrt{1-\frac{n\left( n+1\right) }{j\left(
j+1\right) }},
\end{equation*}
and

\begin{equation*}
\emph{I}_{L}\left( \phi \right) =-\frac{2\chi }{N_{s}-1}\left( 1+\sin
^{2}\phi \right) +
\end{equation*}
\begin{equation*}
\frac{1}{2\pi }\sqrt{j\left( j+1\right) }\cos \phi \sum_{n=-j}^{j}\left(
-1\right) ^{n}\left( \frac{1}{2n+5}+\frac{1}{2n-3}-\frac{2}{2n+1}\right) 
\sqrt{1-\frac{n\left( n+1\right) }{j\left( j+1\right) }}
\end{equation*}
respectively. Furthermore, the remaining summations can be seen to have well
defined limits for $N_{s}\gg 1$, so that we get 
\begin{equation}
\emph{V}_{L}\left( \phi \right) \simeq -\frac{N_{s}+1}{2}\cos \phi -\frac{%
\chi \left( N_{s}+3\right) }{4}\sin ^{2}\phi ,  \label{spinpot}
\end{equation}
and 
\begin{equation}
\emph{I}_{L}\left( \phi \right) \simeq -\frac{2}{N_{s}-1}\cos \phi -\frac{%
2\chi }{N_{s}-1}\left( 1+\sin ^{2}\phi \right)  \label{spininercia}
\end{equation}
respectively. These expressions are closely related to those obtained from
an Adiabatic Time Dependent Hartree-Fock (ATDHF) approach to this model \cite
{holz}. In fact, the expression that gives the inertia function, Eq.(\ref
{spininercia}), almost coincides with the one obtained from that approach ($%
N_{s}$ instead of our $N_{s}-1$) while the expression associated with the
potential, namely 
\begin{equation*}
\emph{V}_{L}^{ATDHF}\left( \phi \right) =-\frac{N_{s}}{2}\left( \cos \phi +%
\frac{\chi }{2}\sin ^{2}\phi \right) ,
\end{equation*}
shows that the ATDHF mean field approximation does not include the term $-%
\frac{1}{2}\cos \phi -\frac{3\chi }{4}\sin ^{2}\phi $ that appears in our
approximation. This difference is due to the initial GCM\ ansatz used in our
approach which deals with linear superpositions of Slater determinants as
the starting quasi-spin wavefunctions instead of pure determinants as it is
the case with ATDHF. A similar discussion using the GCM has been also
presented in Ref. \cite{holz}.

Figure 1 depicts the potential functions obtained from expressions (\ref
{exapot}) and (\ref{spinpot}), while Figure 2 shows the functions given by (%
\ref{exainer}) and (\ref{spininercia}), with the changed signal, for $\chi
=1.5$ and $N_{s}=2,6,10$ and $20$, respectively. From these figures it can
be verified that the curves tend to superpose as $N_{s}$ increases being
that they are almost indistinguishable already for $N_{s}=20$, i.e., $j=10$.

With these results we can now check the validity of our approach and its
limiting version when $N_{s}\gg 1$. To compare the two approaches we first
obtain the exact results by diagonalizing the energy matrix $H_{n,n^{\prime
}}$, Eq.(\ref{exata}), and, on the other hand, by solving the
Schr\"{o}dinger equation 
\begin{equation}
\left[ -\frac{1}{2}\frac{d}{d\phi }\left( \frac{1}{\emph{M}\left( \phi
\right) }\frac{d}{d\phi }\right) +\emph{V}\left( \phi \right) \right] \psi
\left( \phi \right) =E\psi \left( \phi \right)  \label{schrod}
\end{equation}
using the expressions for $\emph{V}\left( \phi \right) $, Eq.(\ref{spinpot}%
), and $\emph{M}\left( \phi \right) $, obtained from Eq.(\ref{spininercia}),
respectively. From the outset it is clear that the second order moment
truncation proposed here will not suffice to describe the quasi-spin system
for small $N_{s}$ since in these cases higher order terms are also
important. Thus the requirement of large $N_{s}$ imposed so that a reliable
two-terms Hamiltonian can be obtained also results in the approximate
expressions, Eqs.(\ref{spinpot}) and (\ref{spininercia}) respectively. In
this form, from a direct inspection of Figures 1 and 2, we know beforehand
that our approach must be acceptable for $N_{s}=2j\gtrsim 10$. This guides
us to study the spectra of systems with those quasi-spin number directly
using expressions (\ref{spinpot}) and (\ref{spininercia}) as our starting
point. Therefore, the solutions thus obtained give an way of testing the
quality of our approximate expressions for the potential and inertia
functions.

A comparison between the approximate and exact energies can be seen in
Figures 3 and 4. A check of our description is shown in Figure 3 where the
exact ground state eigenvalues of the quasi-spin systems, with $%
N_{s}=10,\;20 $ and $40$,$\;$are compared with those obtained from Eq.(\ref
{schrod}), using Eqs.(\ref{spinpot}) and (\ref{spininercia}), as a function
of the strength parameter $\chi $. In Figure 4 we show the energy errors in
the ground state eigenvalues as a function of the quasi-spin number $N_{s}$
for a fixed value of the strength parameter, namely, $\chi =1.0$. The
results are almost the same for other values of $\chi $. It is clear from
Figure 4 that an error as low as $1\%$ is achieved already for a system with 
$N_{s}=8$ $(j=4)$ thus confirming our previous assumption, and therefore
establishing a suitable working limit for the number of quasi-spins of the
system when we use the present approach. This also means that the truncation
of the series of moments of the energy surface up to the second term can,
already for $2j=N_{s}\gtrsim 8$, account for the dominant part of the
dynamical content of the quasi-spin system.

Two interesting features of the Lipkin model can be directly discussed in
this approach. First, it is immediate to see, observing the expression
describing the potential function, Eq.(\ref{spinpot}), that the strength
parameter, namely $\chi $, assumes its critical value $\chi _{c}$ -- the
minimum at $\phi =0$ turns into a maximum -- when $\chi =\chi _{c}=\left(
N_{s}+1\right) /\left( N_{s}+3\right) $. This shows that only when $%
N_{s}\rightarrow \infty $ we get $\chi _{c}\rightarrow 1,$ as expected \cite
{yaffe}. Clearly for values $\chi >\chi _{c}$ $\left( N_{s}<\infty \right) $
a barrier appears at the origin and tunneling can occur under certain
conditions \cite{gapili}. Second, the possibility of tunneling through the
barrier, when it is the case, can be blocked if the effective mass presents
a divergent behaviour at or near the classical turning points, i.e., if $%
I_{L}\left( \phi \right) $ exhibits zeroes there. In this form, not only the
potential barrier governs the tunneling process, but the inertia function
also plays an essential role in this matter. Due to the behaviour of the
inertia some regions in the angle domain may not be acessible; this can also
be seen as a localization of the spin orientation when this singular
behaviour occurs: when it turns out to be impossible for the spin to change
its orientation - when the effective mass diverges - it cannot tunnel.
Considering $I_{L}\left( \phi \right) $ for the Lipkin model we see that its
zeroes occur at $\phi _{0}=\pm \frac{\pi }{2}$ for $\chi =0$, and that they
are pushed to the borders of the angle interval as $\chi $ increases to $1$,
i.e., $\phi _{0}=\pm \pi $ for $\chi =1$. For $\chi >1$ there will be no
zeroes at all. This result shows us that whenever a barrier is present in
the potential function -- and this occurs for $\chi >\chi _{c}$ -- there
will always be a nonvanishing tunneling probability. As a consequence, there
will always be tunneling in the Lipkin model when a barrier is present.

\section{Applications}

\subsection{ Mn-12 Acetate molecule}

If we choose now the parameters of the phenomenological Hamiltonian such
that $A=g\mu _{B}H_{\parallel }$, $B=-D\;(D>0)$, and $G=0$ we have the
already proposed Mn12-acetate molecule Hamiltonian model \cite
{tejada,caneschi}. Starting from that Hamiltonian, and using the general
expression, Eq.(\ref{superficiegeral}), we end up with the corresponding
energy surface 
\begin{equation*}
\emph{H}_{Ac}\left( u,\phi \right) =-\frac{Dj}{4\pi }\sum_{l=-j}^{j}e^{i%
\frac{l}{2}u}-\frac{D}{4\pi }\sum_{l=-j}^{j}e^{i\frac{l}{2}u}\left( j^{2}-%
\frac{l^{2}}{4}\right) -
\end{equation*}
\begin{equation*}
\frac{g\mu _{B}}{2\pi }H_{\parallel }\cos \phi \sum_{l=-j}^{j}e^{i\frac{l}{2}%
u}\sqrt{\left( j+\frac{l}{2}+\frac{1}{2}\right) \left( j-\frac{l}{2}+\frac{1%
}{2}\right) }\;-
\end{equation*}
\begin{equation*}
\frac{D}{4\pi }\cos 2\phi \sum_{l=-j}^{j}e^{i\frac{l}{2}u}\sqrt{\left( j+%
\frac{l}{2}+1\right) \left( j+\frac{l}{2}\right) \left( j-\frac{l}{2}%
+1\right) \left( j-\frac{l}{2}\right) }.
\end{equation*}
We can now obtain the expressions for the potential and inertia functions, 
\begin{equation*}
\emph{V}_{Ac}\left( \phi \right) =-Dj\left( j+1\right) \cos ^{2}\phi -
\end{equation*}
\begin{equation*}
\frac{g\mu _{B}H_{\parallel }}{2j+1}\cos \phi \sqrt{j\left( j+1\right) }%
\sum_{n,k=-j}^{j}e^{2\pi i\frac{k}{N}\left( n+\frac{1}{2}\right) }\sqrt{1-%
\frac{n\left( n+1\right) }{j\left( j+1\right) }}
\end{equation*}
and 
\begin{equation*}
\emph{I}_{ac}\left( \phi \right) =-\frac{1}{M\left( \phi \right) }=-D+\frac{%
g\mu _{B}H_{\parallel }}{4\left( 2j+1\right) }\cos \phi \sqrt{j\left(
j+1\right) }\sum_{n,k=-j}^{j}e^{i\pi \frac{k}{N}}\times
\end{equation*}
\begin{equation*}
\left[ e^{2\pi i\frac{k}{N}\left( n+2\right) }+e^{2\pi i\frac{k}{N}\left(
n-2\right) }-2e^{2\pi i\frac{k}{N}n}\right] \sqrt{1-\frac{n\left( n+1\right) 
}{j\left( j+1\right) }}+
\end{equation*}
\begin{equation*}
\frac{D}{4}\cos 2\phi j\left( j+1\right) \left[ \sqrt{1-\frac{6}{j\left(
j+1\right) }}\sqrt{1-\frac{2}{j\left( j+1\right) }}-1\right]
\end{equation*}
respectively. Since in this case $j=S=10$, and all the summations can be
directly calculated, we are allowed by the estimates of the previous section
to directly use here the already discussed approximations such that these
expressions reduce respectively to 
\begin{equation}
\emph{V}_{Ac}\left( \phi \right) =-S(S+1)D\cos ^{2}\phi -\gamma g\mu
_{B}H_{\parallel }\cos \phi  \label{acepot}
\end{equation}
and

\begin{equation}
\emph{I}_{Ac}\left( \phi \right) =-D-\xi g\mu _{B}H_{\parallel }\cos \phi
-\beta D\cos 2\phi ,  \label{ine}
\end{equation}
where $\gamma \cong \sqrt{S\left( S+1\right) }$ (to within 1.4\%), $\xi
\cong $ $1/S$ (to within 0.1\%), and $\beta \cong 1.0$ (to within 0.47\%).
Figures 5 and 6 depict $\emph{V}_{Ac}\left( \phi \right) $ and $\emph{I}%
_{Ac}\left( \phi \right) $ for some values of $H_{\parallel }$ where we have
adopted the angle domain $\left( -\frac{\pi }{2},\frac{3\pi }{2}\right) $
for the sake of clarity. In Figure 5 we see that the potential for the
Mn12-acetate molecule presents two minima whose energy depths depend on the
applied field. At the same time the expressions for $I_{Ac}\left( \phi
\right) $ shown in Figure 6 present a similar behaviour. As it is shown, as
the paralel field increases, one of the minima gets shallower while the
other deepens as expected from phenomenological considerations \cite
{friedman}.

If we solve the Schr\"{o}dinger equation with the obtained expressions for
the potential and inertia function, we obtain $E_{gs}=-60.278\;K$ for the
ground state energy when $H_{\parallel }=0$ with $D/k_{B}=0.6$ $K$. This
result shows that the calculated ground state energy is only $0.46\%$ below
the exact energy result ($-60.0\;K$), obtained by diagonalizing the
corresponding $H_{n,n^{\prime }}$ matrix. When $H_{\parallel }\neq 0$ the
errors are of the same order of magnitude as that. These results confirm our
previous considerations, and thus allow us to further discuss the
Mn12-acetate molecule properties using our approach. However, instead of
extracting results from numerical calculations based on the new Hamiltonian,
we will discuss some basic properties of this molecule by analysing the
potential and inertia function analytic expressions.

Observing that we can consider $\xi =1/S$, and that we can introduce the
field at which the magnetization of the Mn12-acetate molecule attains
saturation \cite{tejada}, i.e., 
\begin{equation*}
H_{a}=\frac{2SD}{g\mu _{B}}\;,
\end{equation*}
we can further rewrite expressions (\ref{acepot}) and (\ref{ine}) as 
\begin{equation}
\emph{V}_{Ac}\left( \phi \right) =-S\left( S+1\right) D\cos ^{2}\phi -\frac{%
2D}{H_{a}}S\sqrt{S\left( S+1\right) }H_{\parallel }\cos \phi
\label{potacefinal}
\end{equation}
\begin{equation}
\emph{I}_{Ac}\left( \phi \right) =-2D\cos ^{2}\phi -\frac{2D}{H_{a}}%
H_{\parallel }\cos \phi .  \label{ineracefinal}
\end{equation}
It can be immediately seen that while $\emph{V}_{Ac}\left( \phi \right)
\propto S^{2}$, $\emph{I}_{Ac}\left( \phi \right) \propto S^{0}$, thus
showing the relative dominance of the potential energy term in the
approximate Hamiltonian.

Now, let us analyse the expressions for the potential and inertia functions.
First of all let us extract the essential features of the potential
function. It is direct to see that since the treatment is of quantum nature,
it embodies the inherent quantum correlations, and as such, even in the
absence of paralel magnetic fields, the potential minima are $-66.0\;K$,
instead of the semiclassical value $-60.0\;K$ obtained with $D/k_{B}=0.6\;K$%
. In fact, in our calculation, it is the ground state energy that occurs at $%
-60.278\;K$, meaning that this is the energy height of the ground state
barrier in this description, and furthermore that the ground state energy
does not coincide with the bottom of the potential, due to the uncertainty
principle, as it must be. This also points to the fact that the quantum
correction related to the uncertainty principle amounts to about $10\%$ of
the potential barrier height. In the particular case of no paralel field,
the minima occur at $\phi _{\min }=0,\pi $, while the maxima occur at $\phi
_{\max }=\frac{\pi }{2},\frac{3\pi }{2}\left( =-\frac{\pi }{2}\right) $.

The introduction of a paralel magnetic field, $H_{\parallel }$, only shifts
the maxima, the minima being kept unaltered, viz., 
\begin{equation*}
\phi _{\max }=\arccos \left( -\frac{1}{\sqrt{S\left( S+1\right) }}\frac{%
H_{\parallel }}{H_{a}}\right) .
\end{equation*}

In what refers to the inertia function let us study first of all its
expression by looking for the extrema. Let us consider the equation 
\begin{equation*}
\frac{d\emph{I}_{Ac}\left( \phi \right) }{d\phi }=\cos \phi \sin \phi +\frac{%
1}{2}\frac{H_{\parallel }}{H_{a}}\sin \phi =0.
\end{equation*}
As before, when there is no magnetic field, $H_{\parallel }=0$, the maxima
occur at $\phi _{\max }=\pm \frac{\pi }{2},\frac{3\pi }{2}$ respectively,
while the minima occur at $\phi _{\min }=0,\pi $.

Now, it is important to see that the search for the zeroes of the inertia
function gives us the possibility of characterizing, at least qualitatively,
the conditions under which tunneling can occur. Thus, starting from Eq.(\ref
{ineracefinal}), we verify that the zeroes of 
\begin{equation*}
\cos ^{2}\phi +\frac{1}{H_{a}}H_{\parallel }\cos \phi =0
\end{equation*}
will determine the situations for which tunneling will not occur since then
the effective mass goes to infinity, and the spin will be trapped in the
angular region around the minima of the potential. It is then direct to see
that the roots of that equation will be $\phi =\pm \frac{\pi }{2},\frac{3\pi 
}{2}$, and $\phi =\arccos \left( -\frac{H_{\parallel }}{H_{a}}\right) $. The
existence of these zeroes directly imply that tunneling cannot occur;
obviously this conclusion is verified by the exact diagonalization of the
model Hamiltonian for the molecule, i.e., the existence of zeroes in $%
I\left( \phi \right) $ is the way through which the present description
reflects the diagonal behaviour of the phenomenological Hamiltonian in the $%
|Sm\rangle $ basis. In other words, the spectrum then presents a collection
of doubly-degenerate energy levels.

\subsection{ Fe8 cluster}

By choosing now $B=-D$ and $G=E/2$, with $A=0$, we have the proposed
Hamiltonian for the Fe8 cluster \cite{barra,sangregorio,caneschi} in the
absence of external magnetic fields. Thus, by taking into account that,
again, $j=S=10$, we can use all the approximations discussed before so that
we obtain 
\begin{equation}
V_{Fe}\left( \phi \right) =-(D-E)S(S+1)\cos ^{2}\phi -ES(S+1),
\end{equation}
and 
\begin{equation}
I_{Fe}\left( \phi \right) =-\frac{1}{M\left( \phi \right) }=-2\left(
D-E\right) \cos ^{2}\phi -4E,  \label{inefe}
\end{equation}
respectively. In the same way as in the Mn12-acetate, we also have $%
V_{Fe}\left( \phi \right) \propto S^{2},$ and $I_{Fe}\left( \phi \right)
\propto S^{0}$ thus also showing the relative dominance of the potential
energy contribution to the approximate Hamiltonian.\ In Figure (7) we depict
the potential and the effective mass functions for $D/k_{B}=0.275\;K$ and $%
E/k_{B}=0.046\;K$.

The numerical solutions of the Schr\"{o}dinger equation written with those
expressions and constants also give results that agree quite well with the
ones obtained from the diagonalization of the corresponding original model
Hamiltonian. The errors are less than 1\% for the energy values in the lower
part of the spectrum.

As in the Mn12-acetate molecule without paralel magnetic fields, the minima
of the potential occur at\textit{\ }$\phi =0,\pi $\textit{, }while the
maxima occur at\textit{\ }$\phi =\pm \pi /2,3\pi /2$. As such, the energy
height defined by 
\begin{equation*}
V_{Fe}^{\max }\left( \phi =\frac{\pi }{2}\right) -V_{Fe}^{\min }\left( \phi
=0\right) =(D-E)S(S+1)
\end{equation*}
is obviously only dependent on the internal parameters of the molecule.

It is now possible to evaluate the energy barrier height associated with the
lowest energy level. If we consider the energy of the ground state (but not
the botton of the potential well) then, 
\begin{equation*}
h_{b}=-ES(S+1)-E_{gs};
\end{equation*}
using $E_{gs}\cong -27.645\;K$, obtained from the Schr\"{o}dinger equation,
we get $h_{b}\cong 22.58\;K$, which is only $1.7\%$ higher than the
experimental result, $22.2\;K$, presented in \cite{barra}.

In addition, it is important to observe that the top of the potential
barrier lies in the negative energy domain so that there appear some energy
levels -- all those above the barrier -- that seem not to be associated with
tunneling processes. In fact there are some pairs of levels that are almost
degenerate although this is not due to tunneling, as is also the case in the
Lipkin model when $\chi >\chi _{c}$. This effect is due to the particular
form of the Fe8 Hamiltonian which embodies a subtle interplay between the $%
J_{z}^{2}$ term, which tends to produce degenerate doublets, and the $\left(
J_{x}^{2}-J_{y}^{2}\right) =\frac{1}{2}\left( J_{+}^{2}+J_{-}^{2}\right) $
term which tends to symmetrize the energy spectrum about zero.

In what concerns the expression that gives the inertia function, Eq.(\ref
{inefe}), it is immediately seen that it presents its minima precisely at
the same angles as the potential function does. Furthermore, for the
specific values of the Fe8 cluster parameters, the mass function does not
exhibit infinities so that tunneling is always possible, even in the absence
of an external transverse magnetic field. In fact, the tunneling probability
rate is strongly dependent on the behavior of the effective mass function in
the barrier region

\section{Conclusions}

In this contribution we have shown that a Hamiltonian written in terms of an
angle variable can be constructed that can describe spin systems in a
consistent way. The construction scheme starts from a Hamiltonian given in
terms of the operators obeying the angular momentum commutation relations
and that is, in general, introduced from phenomenological considerations.
The quantum Hamiltonian here proposed extends a previous description, that
only took into account a potential energy term, by also introducing a
kinetic energy term in which an effective mass associated with the spin
system is present. Although this Hamiltonian has been obtained in a general
form, the validity of the basic hypothesis supporting the assumed dominance
of the first two terms of a moment expansion of the associated matrix energy
distribution has been tested. Using the soluble Lipkin quasi-spin model as a
testing ground, we have then shown that a criterion for the obtained
two-terms Hamiltonian to be a good description of the system can be
established. From this study it is clear that for systems with $j=S\geq 5$
the energy spectrum - then obtained from a Schr\"{o}dinger equation - is a
very good approximation to the one obtained from the diagonalization of the
starting model Hamiltonian. It is then important to verify that some
recently produced molecules are physical systems of interest that have
global angular momentum values in the range that allow us to use the present
approach. In this perspective, in what concerns the energy spectra and
barriers, we have shown that a consistent description were obtained for the
Mn12-acetate as well as for the Fe8 cluster molecules. Being of quantum
nature, this approach extends the semiclassical estimates and introduce
essential corrections that would not be present.

In what concerns the possibility of spin tunneling, we have shown that in
the present description the potential energy function clearly exhibits
energy barriers whose heigths are in agreement with the experimental
results. At the same time we have emphasized the essential role played by
the effective mass function in the discussion of tunneling in the sense that
its values in the barrier region are a measure of how easy it is for the
system to change its angular orientation. The presence of infinities in the
effective mass function is an unequivocal feature indicating that tunneling
is then forbidden. In the case of the Mn12-acetate molecule the divergences
always occur if only a paralel magnetic field is present, as it should be.
The expression for the effective mass then shows the way through which the
final effective Hamiltonian embodies that well-known basic quantum
mechanical result. The same is not true for the Fe8 cluster when tunneling
can occur, even in the absence of a transverse magnetic field. The
phenomenological Hamiltonian reflects the symmetries of the molecule, and it
is precisely its particular form, with its coefficients obtained from the
experiments, which gives rise to an effective mass function that is nowhere
divergent.

In the present paper we have carried out a qualitative discussion pointing
to the main aspects of our approach that can be used to describe the spin
tunneling process. In a forthcoming contribution we intend to present
quantitative estimates of the energy levels splitting associated with the
spin tunneling.

\begin{acknowledgement}
The author is grateful to Prof. M.A. Novak from Universidade Federal do Rio
de Janeiro, A.F.R. de Toledo Piza from Universidade de S\~{a}o Paulo, and
B.M. Pimentel from Instituto de F\'{i}sica Te\'{o}rica- UNESP for valuable
suggestions. The author was partially supported by Conselho Nacional de
Desenvolvimento Cient\'{i}fico e Tecnol\'{o}gico, CNPq, Brazil.
\end{acknowledgement}


{}\newpage

\bigskip\ \ \ \ \bigskip\ \ \ \ \ \ \ \ \ \ \ \ \ \ \ \ \ \ \ \ \ \ \ \ \ \
\ \ \ \ \ \ \ \ \ \ \ \ \ \ \ \ \textbf{Figure Captions}

Figure 1: Comparison between the large $N_{s}$ potential function
(continuous line) and the full expression, Eq.(8), (dashed line) for the
Lipkin model for different values of $N_{s}$. The interaction strength is $%
\chi =1.5$ in all cases.

\medskip

Figure 2: Comparison between the large $N_{s}$ function $-I(\phi)$
(continuous line) and the full expression, Eq.(9), (dashed line) for the
Lipkin model for different values of $N_{s}$. The interaction strength is $%
\chi =1.5$ in all cases.

\medskip

Figure 3: Relative error in the ground state energies obtained from Eq.(\ref
{schrod}) for the Lipkin model when compared with the exact results, as a
function of the interaction strength $\chi $. The continuous curve is
associated with $j=5\;(N_{s}=10)$, the dashed curve with $j=10\;\left(
N_{s}=20\right) $, and the dot-dashed curve with $j=20\;(N_{s}=40)$. The
lines only guide the eyes.

\medskip

Figure 4: Relative error in the ground state energies obtained from Eq.(\ref
{schrod}) for the Lipkin model, when compared with the exact results, as a
function of the number of spins, $N_{s}$\ $\left( =2j\right) $, for $\chi
=1.0$. For systems with $N_{s}>10$ the deviation lies below $1\%$. The line
only guides the eyes.

\medskip

Figure 5: Curves representing the potential energy of the Mn12-acetate for
some different values of the paralel magnetic field.

\medskip

Figure 6: Curves representing the function $I\left( \phi \right) $ of the
Mn12-acetate for some different values of the paralel magnetic field.

\medskip

Figure 7: Potential and effective mass functions associated with the Fe8
cluster.

\end{document}